\documentclass[11pt]{article}
\usepackage{amssymb}
\usepackage{amsmath}
\usepackage{amsthm}

\usepackage{graphicx,color}

\newcommand*{\cE}{\mathcal{E}}
\newcommand*{\cL}{\mathcal{L}}

\def\openone{\leavevmode\hbox{\small1\kern-3.8pt\normalsize1}}

\newtheorem{proposition}{Proposition}

\theoremstyle{definition}

\newcommand{\pder}[2]{\frac{\partial #1}{\partial #2 }}

\begin{document}
\title{\LARGE\bf
 Entropy, subentropy and\\ the elementary symmetric functions}
\author{Richard Jozsa
  and Graeme Mitchison\\[3mm]
  \small\it
  \small\it DAMTP, Centre for Mathematical Sciences, University of Cambridge,\\ \small\it Wilberforce Road, Cambridge CB3 0WA, U.K.}

\date{}

\maketitle

\begin{abstract}
We use complex contour integral techniques to study the entropy $H$ and subentropy $Q$ as functions of the elementary symmetric polynomials, revealing a series of striking properties. In particular for these variables, derivatives of $-Q$ are equal to derivatives of $H$ of one higher order and the first derivatives of $H$ and $Q$ are seen to be completely monotone functions. It then follows that $\exp (-H)$ and $\exp( -Q)$ are Laplace transforms of infinitely divisible probability distributions.
\end{abstract}

It is a striking fact that the entropy H and subentropy Q (cf. \cite{ddbj,jrw}) are symmetric functions of their arguments:
\begin{equation}\label{heq} H(x_1, \ldots ,x_d) =-\sum_{i=1}^d x_i \ln x_i \end{equation}
\begin{equation} \label{qdeq}
Q(x_1, \ldots ,x_d)= - \large\sum_{i=1}^d  \frac{x_i^d}{ \prod_{ j \ne i}(x_i - x_j)}\, \ln x_i .
\end{equation}
It is thus perhaps natural to study them as functions of the associated elementary symmetric polynomials defined by
\[ e_1=\sum_j x_j, \hspace{5mm} e_2=\sum_{i<j} x_ix_j, \hspace{5mm} e_3=\sum_{i<j<k} x_ix_jx_k, \hspace{5mm}\ldots  \]
Here we will lift the probability condition $e_1=\sum x_j =1$ and view $e_1, e_2, \ldots , e_d$ as independent variables. Without loss of generality we will list the $x_j$'s in non-increasing order $0<x_1\leq x_2 \leq \ldots \leq x_d$.

For the case of $H$ Fannes \cite{fannes} recently obtained the following elegant expression for the derivatives of H with respect to $e_2,e_3, \ldots ,e_d$ (cf. eq. (9) of \cite{fannes}):
\begin{equation}\label{der}
\frac{\partial H}{\partial e_k}= \int_0^\infty dt \frac{t^{d-k}}{(t^d+e_1t^{d-1} +e_2t^{d-2} +\ldots +e_d)}
\hspace{5mm} k=2,\ldots ,d, \end{equation}
and in particular deduced that $\partial H/\partial e_k \geq 0$ for $k\geq 2$ (which had been shown previously by other means in \cite{mj04}).  Fannes established  eq. (\ref{der}) by starting with an inscrutably ingenious integral identity (eq. (6) in \cite{fannes}). Here we will give an alternative derivation based on complex contour integration techniques and we will also treat the case of subentropy as a function of the elementary symmetric polynomials. Our formulae will reveal a remarkable relationship between the derivatives of entropy and subentropy viz.
\[  -\pder{Q}{e_k}=\frac{\partial^2 H}{\partial e_l \,\partial e_m} \hspace{3mm}\mbox{for any $k,l,m$ with $k=l+m$ and $l,m\geq 1$.} \]
We will also point out a series of further properties of (higher order) derivatives of $H$ that follow directly from eq. (\ref{der}) (and also from our contour integral expressions) and which establish the property of complete monotonicity of $\partial H / \partial e_k$ for $k\geq 2$ on $\{ (e_1, \ldots ,e_d): \mbox{$e_k>0$ for all $k$}\}$.

We begin with the fundamental relation between the $e_k$'s and $x_j$'s viz. that $x_1, \ldots ,x_d$ are the roots of the polynomial equation
\begin{equation}\label{star} 
x^d -e_1 x^{d-1}+e_2x^{d-2} - \ldots +(-1)^d e_d=0. \end{equation}
This defines each $x_j$ implicitly as a function of the $e_k$'s and implicit differentiation gives 
\[ \pder{x_j}{e_k}=\frac{(-1)^{k+1} x_j^{d-k}}{\prod_{i\neq j} (x_j-x_i)} \]
so then the chain rule gives (as elaborated in \cite{mj04}  eqs. (10) - (16))
\begin{equation}\label{derk}
\pder{H}{e_k}=(-1)^k \sum_{j=1}^d \frac{x_j^{d-k}\ln x_j}{\prod_{i\neq j} (x_j-x_i)} \hspace{3mm} \mbox{for $k\geq 2$} \end{equation}
and
\begin{equation}\label{der1}
\pder{H}{e_1}=-\sum_{j=1}^d \frac{x_j^{d-1}\ln x_j}{\prod_{i\neq j} (x_j-x_i)} -1 \hspace{3mm} \mbox{for $k=1$} 
\end{equation}
Next note that by Cauchy's integral formula we have, for any holomorphic function $g$,
\begin{equation}\label{oint}
\sum_{j=1}^d \frac{g(x_j)}{\prod_{i\neq j} (x_j-x_i)}= \frac{1}{2\pi i} \oint \frac{g(z)}{(z-x_1)\ldots (z-x_d)} \, dz \end{equation}
where the contour surrounds all poles at $z=x_1, \ldots ,x_d$ and $g$ is holomorphic in and on the contour. Then eqs. ({\ref{derk}) and (\ref{der1}) immediately give
\begin{equation}\label{ointk}
\pder{H}{e_k}= \frac{1}{2\pi i} \oint \frac{(-1)^k z^{d-k}\ln z}{(z^d-e_1z^{d-1} + \ldots +(-1)^de_d)}\, dz
\hspace{3mm}\mbox{for $k\geq 2$} \end{equation}
and
\begin{equation}\label{oint1}
\pder{H}{e_1}= \frac{1}{2\pi i} \oint \frac{-z^{d-1}\ln z}{(z^d-e_1z^{d-1} + \ldots +(-1)^de_d)}\, dz\,\, -1
\hspace{3mm}\mbox{for $k=1$} .\end{equation}
In all these cases the contour goes around all $0<x_1\leq x_2 \leq \ldots \leq x_d$ on the real $z$-axis but not around the branch point $z=0$ of $\ln z$.

Now to regain Fannes' formula eq. (\ref{der}) we distort the contour into a keyhole contour that excludes the negative real $z$-axis i.e. it runs above and below the negative real axis at distance $\epsilon$ between $z=-R\pm i\epsilon$ and $z=0\pm i\epsilon$, loops around the origin $z=0$, and is completed by a circle of (large) radius $R$. Then direct calculation using standard contour integration techniques (cf. \cite{cplexanalysis}) with the limits $\epsilon \rightarrow 0$ and $R\rightarrow \infty$ gives Fannes' formula for the case of $k\geq 2$.

The case of subentropy is easier since $Q$ itself is already of the form of the LHS of eq. (\ref{oint}) and we immediately get (with the same contour as used above):
\begin{equation}\label{k3}
Q=-\frac{1}{2\pi i} \oint \frac{z^{d}\ln z}{(z^d-e_1z^{d-1} + \ldots +(-1)^de_d)}\, dz.
\end{equation}
By looking at eqs. (\ref{ointk}), (\ref{oint1}) and (\ref{k3}) we easily see the following relation.
\begin{proposition}\label{prop1}
\[  -\pder{Q}{e_k}=\frac{\partial^2 H}{\partial e_l \,\partial e_m} \hspace{3mm}\mbox{for any $k,l,m$ with $k=l+m$ and $l,m\geq 1$.} \,\,\, \Box\]
\end{proposition}
Returning now to eq. (\ref{der}) it is easy to similarly see that higher derivatives of $H$ with respect to the $e_k$'s satisfy the properties in the following three propositions.
\begin{proposition}\label{prop2} For $m\geq 2$ we have
\[ (-1)^{m-1} \frac{\partial^m H}{\partial e_{i_1}\ldots \partial e_{i_m}} \geq 0\hspace{3mm} \mbox{for all $i_1,\ldots ,i_m \geq 1$,}\]
and for $m=1$ we have
\[ \pder{H}{e_k} \geq 0 \hspace{3mm} \mbox{for $k\geq 2$.}\,\,\, \Box \]
\end{proposition}
\begin{proposition}\label{prop3}
The $m^{\rm th}$ derivative \[ \frac{\partial^m H}{\partial e_{i_1}\ldots \partial e_{i_m}}\hspace{3mm} \mbox{ as a function of $(e_1, \ldots e_d )$} \] depends only on the sum of indices $i_1+\ldots +i_m$, and the same property holds for $Q(e_1, \ldots ,e_d)$ too.\, $\Box$
\end{proposition}
\noindent Thus for example $\partial^2 H/\partial e_1 \partial e_5 = \partial^2 H/\partial e_2 \partial e_4=\partial^2 H/\partial e_3^2$ since $1+5=2+4=3+3$.

Some of the above formulae appear to become singular if any of the $x_j$'s coincide (e.g. if $x_1=x_2$). However closer inspection reveals that the limit of coincidence (e.g. $x_1\rightarrow x_2$) is always finite and in the contour integral formulae we just use Cauchy's integral formula with higher order poles to provide values of derivatives rather than values of the functions themselves. With this in mind we have the following result.
\begin{proposition}\label{prop4} Consider the $m^{\rm th}$ derivative $\frac{\partial^m H}{\partial e_{i_1}\ldots \partial e_{i_m}}$ for $H(e_1, \ldots ,e_d)$ with $d$ variables. Introduce the entropy function $\tilde{H}(\tilde{x}_1, \ldots , \tilde{x}_{dm})$ with $dm$ variables and corresponding elementary symmetric polynomials $\tilde{e}_1, \ldots , \tilde{e}_{dm}$. Then for any $(e_1, \ldots , e_d)$ arising from roots $x_1, \ldots x_d$ we have
\begin{equation}\label{derm1} (-1)^{m-1}\frac{\partial^m H}{\partial e_{i_1} \ldots \partial e_{i_m}} (e_1, \ldots ,e_d) =
\pder{\tilde{H}}{\tilde{e}_{K}}(\tilde{e}_1, \ldots , \tilde{e}_{md}) \end{equation}
where $K=i_1+\ldots +i_m$ and the RHS is evaluated at the point $(\tilde{e}_1, \ldots , \tilde{e}_{md})$ being the elementary symmetric polynomial values for the $md$  $\tilde{x}_j$'s 
\[  (\tilde{x}_1, \ldots , \tilde{x}_{dm}) = (x_1, \ldots , x_1, x_2, \ldots ,x_2, \hspace{2mm} \ldots\hspace{2mm} , x_d, \ldots ,x_d)\]
having each $x_i$ repeated $m$ times.\, $\Box$
\end{proposition}
\noindent {\bf Proof}\, By factoring $(z^d-e_1z^{d-1} +\ldots +(-1)^de_d)$ as $(z-x_1)\ldots (z-x_d)$ we see that 
\[ (z^d-e_1z^{d-1} +\ldots +(-1)^de_d)^m = z^{md}-\tilde{e}_1z^{md-1}+
\ldots +(-1)^{md}\tilde{e}_{md} \] where the $\tilde{e}_k$'s are the
elementary symmetric functions of $md$ variables evaluated at the
repeated values of the $x_j$'s. Then eq. (\ref{derm1}) follows by
differentiating eqs. (\ref{ointk}) and (\ref{oint1}) $m-1$ times.\,
$\Box$

To conclude, we make a connection with the concept of complete
monotonicity and the classical theorem of Bernstein. In \cite{fannes2}
it was shown that these concepts apply to a special kind of entropy;
here we show how they relate to $H$ and $Q$.

A function $f(t_1, \ldots, t_m)$ is said to be {\em completely
  monotone} if
\begin{align}
\label{cm}(-1)^{j}\frac{\partial}{\partial t_{i_1}} \ldots
  \frac{\partial}{\partial t_{i_j}} \ f \ge 0,
\end{align}
for $t_{i_q} \in [0,\infty)$ and $j=0,1,2 \ldots$. From Proposition
\ref{prop2} it follows that each first derivative $\partial H/\partial
e_k$, $2 \le k \le d$, is completely monotone in the variables $e_1,e_2,
\ldots, e_d$, and, using Proposition \ref{prop1}, the same holds for
the derivatives $\partial Q/\partial e_k$.

Bernstein's theorem \cite{feller}, in a multivariate form, says that
any completely monotone function is the Laplace transform of a
positive density, $f(t_1, \ldots, t_m)=\cL[\mu(s_1, \ldots, s_m)](t_1,
\ldots, t_m)$, or more explicitly
\[
f(t_1, \ldots, t_m)=\int_0^\infty e^{-(t_1s_1 + \ldots +t_ms_m)} \mu(s_1, \ldots ,s_m) ds_1 \ldots ds_m.
\]
We can immediately apply this theorem to the derivatives of $H$ and
$Q$. Let us assume $e_1=1$ henceforth. Note that complete
monotonicity and Berstein's theorem require us to consider all of the
positive cone $\cE^+$ defined by $e_k \ge 0$, $2 \le k \le d$, though
only part of this cone corresponds to probabilities, i.e. to real,
positive $x_i$.  For instance, for $d=2$ the roots of $z^2-z+e_2=0$
are real and positive if and only if $e_2 >1/4$, and for larger $d$
there are polynomial conditions on the $e_k$. To obtain the whole of
$\cE^+$ we need to include complex conjugate pairs of roots as well as
positive real roots. However, these complications need not trouble us
when viewing $H$, $Q$ and their derivatives within $\cE^+$.

Of course, the real objects of interest are $H$ and $Q$ themselves
rather than their first derivatives. What can we say about the signs
of $H$ and $Q$? We know $H=0$ when $e_1=1$ and $e_2= \ldots =e_d=0$,
since this corresponds to one of the underlying probabilities being
one and the others zero. But then every point in the positive cone
$\cE^+$ defined by $e_k \ge 0$, $2 \le k \le d$, can be reached by
moving along its coordinate axes independently, and it follows from
$\partial H/\partial e_k \ge 0$ that $H$ must be positive everywhere
in $\cE^+$. Using Proposition \ref{prop1}, a similar conclusion
applies to $Q$.

Thus $H$ itself is not completely monotone, since both $H$ and its
derivatives are positive: there is no change of sign between the
function and its first derivative, as eq. (\ref{cm})
requires. However, if the first derivatives of a function $f$ are
completely monotone, then so is $e^{-f}$ \cite{feller}. This is easy
to check by repeated differentiation of $e^{-f}$. Extending this to
many variables, we see that $e^{-H(e_2, \ldots, e_d)}$ is the Laplace
transform of a completely positive function $\mu(s_2, \ldots, s_d)$,
and since $H(0, \ldots, 0)=0$, $\mu$ is a probability
density. Actually, we can say more than this, since
$e^{-H}=(e^{-H/m})^m=\cL[\nu^{*m}]$, where $e^{-H/m}=\cL[\nu]$. This
means that, for any integer $m$, $\mu$ is the $m$-fold convolution of
a measure $\nu$. This property is called {\em infinitely divisibility}
\cite{feller}, and is possessed by many fundamental
statistical distributions, like the Gaussian.

Thus we know that $e^{-H}$ is the Laplace transform of an infinitely
divisible function, and, since all the above remarks apply to $Q$, the
same is true of $e^{-Q}$. It would be very desirable to be able to
identify these fundamental-seeming underlying
distributions. Unfortunately, we have so far been unable to derive
them, even for $d=2$, and we offer it as an intriguing unsolved
problem.

\

\end{document}